\documentclass[a4paper,twocolumn,preprintnumbers,amsmath,amssymb]{revtex4}

\usepackage{graphicx}
\usepackage{dcolumn}
\usepackage{bm}

\begin{document}
\newcommand{\braopket}[3]{\langle #1 | \hat #2 |#3\rangle}
\newcommand{\braket}[2]{\langle #1|#2\rangle}
\newcommand{\bra}[1]{\langle #1|}
\newcommand{\braketbraket}[4]{\langle #1|#2\rangle\langle #3|#4\rangle}
\newcommand{\braop}[2]{\langle #1| \hat #2}
\newcommand{\ket}[1]{|#1 \rangle}
\newcommand{\ketbra}[2]{|#1\rangle \langle #2|}
\newcommand{\op}[1]{\hat {#1}}
\newcommand{\opket}[2]{\hat #1 | #2 \rangle}

\title{Upper bound for the conductivity of nanotube networks }

\author{L. F. C. Pereira$^{(a)}$, C. G. Rocha$^{(b)}$, A. Latg\'e$^{(c)}$, J. N. Coleman$^{(a)}$ and M. S. Ferreira$^{(a)}$}
\affiliation{
(a) School of Physics and CRANN, Trinity College Dublin, Dublin 2, Ireland \\
(b) Institute for Materials Science and Max Bergmann Center of Biomaterials, Dresden University of Technology, D-01062 Dresden, Germany \\
(c) Instituto de F\'{\i}sica, Universidade Federal Fluminense, Niter\'oi, Brazil  }

\date{\today}

\begin{abstract}

Films composed of nanotube networks have their conductivities regulated by the junction resistances formed between tubes. Conductivity values are enhanced by lower junction resistances but should reach a maximum that is limited by the network morphology. By considering ideal ballistic-like contacts between nanotubes we use the Kubo formalism to calculate the upper bound for the conductivity of such films and show how it depends on the nanotube concentration as well as on their aspect ratio. Highest measured conductivities reported so far are approaching this limiting value, suggesting that further progress lies with nanowires other than nanotubes.

\end{abstract}

\maketitle
\bibliographystyle{apsrev} 

The search for thin films that are flexible, transparent and conductive is driven by their potential as transparent electrodes \cite{Nomura:04, Lewis:04, Jackson:05}. One common route is to use films made by a disordered network of carbon nanotubes (NT) \cite{Snow:03, Wu:04, Gruner:05, Takenobu:06, Martel:06, Rogers:08}. In this case, electrons move across the entire film by moving between NT in close proximity. The conductivity is limited by the tunneling between tubes, which introduces a significant inter-NT junction resistance. To make the films more conductive one needs to improve the coupling between NT, which recently has been achieved with acid treatments \cite{Geng:07, johnny}. Further attempts are being made to lower the junction resistance and surpass the best conductivity reported so far, which currently stands at $\sigma \approx 6 \times 10^5 \, {\rm S/m}$ \cite{Wu:04, Geng:07, johnny}.

In addition to the junction resistance, the network morphology also plays a role in limiting the film conductivity. In fact, we have recently demonstrated how sensitive to the network connectivity the conductivity can be \cite{jap-2008}. This means that no matter how much progress is made in lowering the junction resistance, there should be a maximum value for the film conductivity, which is regulated by the network itself.  This is the goal of the present manuscript, {\it i.e.}, to obtain an upper bound for the conductivity of disordered NT-networks. The knowledge of this upper bound should avoid overoptimistic expectations for the transport properties of the films. Furthermore, understanding the interplay between network morphology and the intrinsic conductances of NT may be explored to deal with films made of other nanowires (NW) \cite{nw, nw2}.   

Since we are interested in the best-case scenario in which the electronic conductivity is at its maximum, we must eliminate potential sources of scattering and decoherence such as structural imperfections, impurities and interaction with other quasi-particles. In this situation it is appropriate to consider a purely ballistic regime of transport within the NW, which calls for a quantum description of the conductivity. NT are known to behave as ballistic conductors with two quanta of conductance \cite{Charlier:07} and are often referred to as possessing two conducting channels. How much interference there is between these channels is what determines how the network affects the film conductivity. Furthermore, because other quantum wires are populated by a different number of conducting channels, the results here obtained are not exclusive for NT and should be also applicable to other materials. With this generality in mind, we introduce a model that describes the transport properties of a network of quantum wires, each one of them capable of carrying $M$ channels of conductance. The model consists of two parts, one macroscopic in which geometrical constraints define the connectivity of the network, and another of microscopic origin that accounts for the electronic structure of the wires. 

We start with the macroscopic part of the model by assuming that a film in contact to two electrodes is represented by a number of rods of length $\ell$ and diameter $d$ randomly distributed inside a rectangular box of dimensions $L \times L \times 2L$. The opposing square faces of the box are taken as the electrodes. It is a simple task to count the number of rods $\langle \alpha_E \rangle$ crossing the electrode walls as well as  $\langle \alpha \rangle$, here defined as the average number of contacts (per rod) between neighboring rods. 
Geometrical arguments  \cite{jap-2008} not involving any wire-specific information other than their aspect ratio ($\ell /d$) can be used to show that the latter quantity  scales linearly with $V_f (\ell / d)$, $V_f$ being the volume fraction of the network, whereas $\langle \alpha_E \rangle$ scales with $V_f (\ell / d)^2$. The universality in these scaling laws means that films composed of different values of $V_f$, $\ell$ and $d$ may have similar connectivities. 

For the sake of generality and simplicity we choose to represent individual wires by 1-dimensional atomic chains within the tight-binding model with M-fold degeneracy in the orbital degrees of freedom. In this way, the Hamiltonian ${\hat h}_j$ associated with a single wire labelled $j$ corresponds to a linear chain that carries $M$ quanta of conductance, {\it i.e.}, ${\hat h}_j = \sum_{n,n^\prime,\mu} \vert n, j, \mu\rangle t \langle n^\prime, j, \mu \vert$, where the state $\vert n, j, \mu\rangle$ represents an atomic orbital $\mu$ localized at an atom numbered $n$ within wire $j$ and the sum over $n$ and $n^\prime$ is for nearest neighbors only. The sum over $\mu$ ranges from 1 to M, which guarantees its M-fold degeneracy, and the parameter $t$ defines the bandwidth of the individual wire. As we shall see, other Hamiltonians could have been chosen without major changes to our conclusions.

The Hamiltonian ${\hat H}$ associated with the network is not a mere sum of the individual ${\hat h}_j$ over the index $j$ but it is dependent on how they are connected to each other. To assemble the Hamiltonian of the full network we assume a number $N$ of finite-sized atomic chains, each one of which containing $1.6 \times 10^4$ atomic sites \cite{obs}. Bearing in mind that these chains ultimately represent NW of length $\ell$ and diameter $d$, we can easily obtain the volume fraction $V_f$ and the corresponding values for $\langle \alpha \rangle$ and $\langle \alpha_E \rangle$. The network is then created as follows. Chains are connected randomly so that there is a total of $N \langle \alpha \rangle / 2$ connections. The intra-chain atomic sites that make the connections are also randomly chosen. The connection is introduced by a hopping probability $\gamma$ inserted in the Hamiltonian connecting two different chains, that is, by a potential  ${\hat V} = \vert n, j, \mu \rangle \, \gamma \, \langle n^\prime, {j^\prime}, \mu \vert$. Attention is given to the fact that $j$ and $j^\prime$ must necessarily be different to represent distinct wires. Notice that this contribution is diagonal in $\mu$ indicating that the inter-wire coupling does not mix the $M$ conducting channels carried by each wire. Regarding the role played by the electrodes, they act as charge reservoirs and can be mimicked by using semi-infinite chains. Since $\langle \alpha_E \rangle$ gives the average number of contacts to the electrodes (per unit area), we must include the appropriate number of semi-infinite chains $N_E = 2 \langle \alpha_E \rangle$ into the network. In summary, our model consists of a network composed of $N$ finite chains randomly connected to each other plus $N_E$ semi-infinite ones, each of which connected to a single finite-sized counterpart. The semi-infinite chains can be sub-grouped into two sets representing contacts to the left and right electrodes. Furthermore, no finite-length chains are allowed to be connected to more than a single semi-infinite one. This requirement is necessary to avoid short circuits.

Having defined the network Hamiltonian, we can now use the Kubo formalism, to calculate the zero-bias conductance $\Gamma$ across the film \cite{datta-book, details, mathon}. Absolutely equivalent to the Landauer formalism, the Kubo formula for the conductance involves only a few Green-function (GF) matrix elements evaluated at the Fermi energy, which can be easily obtained by efficient inversion techniques \cite{ultrasonics}. One of the advantages of using this formalism is that any changes in the Hamiltonian are automatically accounted for in the corresponding GF. As we shall see, replacing the linear-chain Hamiltonian with another that  describes the detailed atomic structure of NT becomes a straightforward task, albeit more computationally demanding. The coupling between wires is regulated by the parameter $\gamma$. Since we are interested in the upper bound for the conductivity, we can increase $\gamma$ until the conductivity saturates. At that point we can say, by inspection, that the conductance across any two wires connected by this value of $\gamma$ matches the intrinsic conductance of the wires, {\it i.e}, $M \Gamma_0$, where $\Gamma_0$ is the quantum of conductance. Inevitably, there will be some degree of reduction on the conductance when two NW are connected, particularly in the case of finite sized wires but the best way to minimize this reduction is to consider very large wires. In the case of atomic chains, $\gamma = t$ is an excellent choice for the coupling parameter since it reduces the conductance between two chains only by a very small fraction. 

With all the parameters defined, the upper bound for the network conductance can now be calculated for different volume fractions and aspect ratios. Each calculation involves a large configurational average in order to achieve statistical significance. Rather than the calculated conductance values, we present in Fig. 1 the respective conductivities plotted as a function of $x \equiv V_f \, \ell / d^2$. With NT in mind, the number of conducting channels is $M = 2$. A wide range of lengths, diameters and volume fractions have been considered but, because of the universality behavior of the connectivities, they all fall onto a common straight line (square symbols) described by $\sigma_u = \beta_M \, x$, where $\beta_M = M \times 4.25 \times 10^{-5} \, {\rm S}$. This indicates that the dominant factor in determining the conductivity of the network is clearly the number of connections with the electrodes, which also scales linearly with $x$. 

\begin{figure}
\includegraphics[width = 8cm] {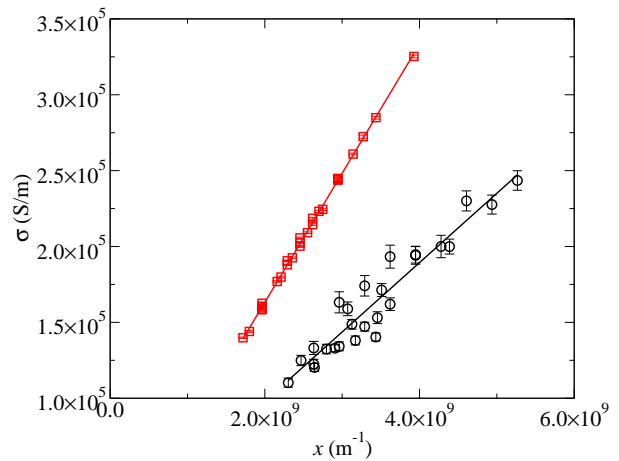}
\caption{Upper bound for the conductivity $\sigma_u$ (in units of ${\rm S/m}$) as a function of the quantity $x = V_f \ell /d^2$ (in units of ${\rm m^{-1}}$). Line with square symbols is for an array of linear chains, each one of which capable of carrying $2 \Gamma_0$ of conductance. Circles correspond to results for disordered arrays of finite-sized NT in which the details of their electronic structure have been fully taken into consideration. In this case, $\gamma =  \gamma_0$, where $\gamma_0$ is the tight-binding electronic hopping between nearest-neighbor carbon atoms of the NT. Error bars account for the standard error in the configurational average.}
\label{figure_1}
\end{figure}

The expression above provides the upper bound for the conductivity of disordered networks in the case of wires capable of carrying $M$ quanta of conductance. It is instructive to test the expression for typical NT values, namely $\ell = 1 \mu {\rm m}$ and $d = 1.2 \, {\rm nm}$. For $V_f = 30\%$, the predicted upper bound would be $\sigma_u = 1.8 \times 10^7 {\rm S/m}$ if NT of these dimensions could be fully dispersed to form the network. NT are however known to bundle together, which means that in reality wire diameters are considerably larger. On the other hand, larger-diameter bundles have more NT on the surface leading to more current-carrying channels per wire. Taking all this into consideration, we can compare our expression with the highest-conductivity case reported so far ($V_f=30\%, \ell = 5 \, {\rm \mu m}, d = 20 \, {\rm nm}$) \cite{Wu:04, Geng:07, johnny, evelyn}. Our prediction of $\sigma_u = 9 \times 10^6 \, {\rm S/m}$ is only one order of magnitude superior to the measured value of $\sigma = 6 \times 10^5 \, {\rm S/m}$. Bearing in mind that the upper bound here obtained assumes a number of ideal conditions that are experimentally unavoidable, this might be a clear indication that we are approaching a saturation point in the conductivity of NT-network films. 

No qualitative change is observed when the linear-chain Hamiltonians  ${\hat h}_j$ describing the individual NW is replaced with another that accounts for the precise electronic structure of NT. In this case, a straight line with circular symbols depicts the conductivity of a network of realistic NT. The difference in slope and the fact that individual values are somewhat more staggered than the linear-chain results is easily explained by finite-size effects. In the case of ${\hat h}_j$ describing real NT, the number of atoms required to generate a wire of similar length $\ell$ is considerably larger. This introduces undesirable fluctuations in the conductance of individual NT, which on average lowers the overall conductivity. By making the wires longer, the fluctuations are reduced and the results for NT tend to approach those for the linear chain. While we cannot increase the NT size without paying hefty computational penalties, we can reduce the size of the liner chains for the sake of comparison. In doing so, the linear behavior observed for linear-chain Hamiltonians coincides with that for NTs. This suggests that the line with square symbols is therefore the most representative for the upper bound for the conductivity of a NT network. 

Finally, although our focus has been on disordered networks comprised of NT, we can extend our results to deal with other wires. This could represent the case of networks made of other conducting materials, such as noble-metal wires, for instance. In this case the number of conducting channels M depends linearly on the wire diameter and the overall conductivity of the network is likely to scale inversely with $d$, with a proportionality constant that depends on the specifics of the wire in question. If the conductivity of NT-network films is approaching its saturation point, it is likely that wires other than NT may occupy the post of ideal components for disordered-network films.   

In summary, we have calculated the upper bound for the conductivity of NT-network films by assuming ideal contacts between tubes. Our results may be used to indicate how much room there is to lower the junction resistance within a film. More importantly, when compared with the highest measurements reported for NT-network films, the upper bound presented here points to a situation in which the conductivity is approaching its limiting value.


\begin{references}
\bibitem{obs} The chain size was chosen to minimize fluctuations in the local density of states.
\bibitem{details} Details of the calculational procedure will be published elsewhere.
\end{references}
\end{document}